\documentclass[prl,twocolumn,preprintnumbers,amsmath,amssymb,superscriptaddress]{revtex4} 

\usepackage{graphicx}
\usepackage{dcolumn}
\usepackage{bm}

\begin{document}

\preprint{}

\title{Coherent phonon manipulation in coupled mechanical resonators}

\author{Hajime Okamoto}
 \email{okamoto.hajime@lab.ntt.co.jp}
\affiliation{NTT Basic Research Laboratories, Nippon Telegraph and Telephone Corporation, Atsugi 243-0198, Japan}
\author{Adrien Gourgout}
\affiliation{NTT Basic Research Laboratories, Nippon Telegraph and Telephone Corporation, Atsugi 243-0198, Japan}
\author{Chia-Yuan Chang}
\affiliation{NTT Basic Research Laboratories, Nippon Telegraph and Telephone Corporation, Atsugi 243-0198, Japan}
\affiliation{The Department of Materials Science and Engineering, National Chiao-Tung University, Hsinchu 980-8578, ROC}
\author{Koji Onomitsu}
\affiliation{NTT Basic Research Laboratories, Nippon Telegraph and Telephone Corporation, Atsugi 243-0198, Japan}
\author{Imran Mahboob}
\affiliation{NTT Basic Research Laboratories, Nippon Telegraph and Telephone Corporation, Atsugi 243-0198, Japan}
\author{Edward Yi Chang}
\affiliation{The Department of Materials Science and Engineering, National Chiao-Tung University, Hsinchu 980-8578, ROC}
\author{Hiroshi Yamaguchi}
\affiliation{NTT Basic Research Laboratories, Nippon Telegraph and Telephone Corporation, Atsugi 243-0198, Japan}


%\begin{abstract}
%
%\end{abstract}

%\pacs{}

\maketitle

{\bf Coupled mechanical oscillations were first observed in paired pendulum clocks in the mid-seventeenth century and were extensively studied for their novel sympathetic oscillation dynamics\cite{Huygens,Pikovsky}. In this era of nanotechnologies, coupled oscillations have again emerged as subjects of interest when realized in nanomechanical resonators for both practical applications and fundamental studies\cite{Spletzer,Gil-Santos,OkamotoAPL,Bannon,KarabalinPRL,Shim,OkamotoAPEX,Sun,Kotthaus}. However, a key obstacle to the further development of this architecture is the ability to coherently manipulate the coupled oscillations. This limitation arises as a consequence of the usually weak coupling between the constituent nanomechanical elements. Here, we report parametrically coupled mechanical resonators in which the coupling strength can be dynamically adjusted by modulating (pumping) the stress in the mechanical elements via a piezoelectric transducer. The parametric control enables the coupling rate between the two resonators to be made so strong that it exceeds their intrinsic energy dissipation rate by more than a factor of four. This ultra-strong coupling can be exploited to coherently transfer phonon populations, namely phonon Rabi oscillations\cite{O'Connell,KippenbergNATURE}, between the mechanical resonators via two coupled vibration modes, realizing superposition states of the two modes and their time-domain control. More unexpectedly, the nature of the parametric coupling can also be tuned from a linear first-order interaction to a non-linear higher-order process in which more than one pump phonon mediates the coherent oscillations. This demonstration of multi-pump phonon mixing echoes multi-wave photon mixing\cite{KippenbergScience} and suggests that concepts from non-linear optics can also be applied to mechanical systems. Ultimately, the parametric pumping is not only useful for controlling classical oscillations\cite{Yamaguchi} but can also be extended to the quantum regime\cite{O'Connell,KippenbergNATURE,TeufelGround,Painter,Palomaki}, opening up the prospect of entangling two distinct macroscopic mechanical objects\cite{Wineland,Harlander}.}

\begin{figure*}[tbp]
\begin{center}
\includegraphics[scale=0.47]{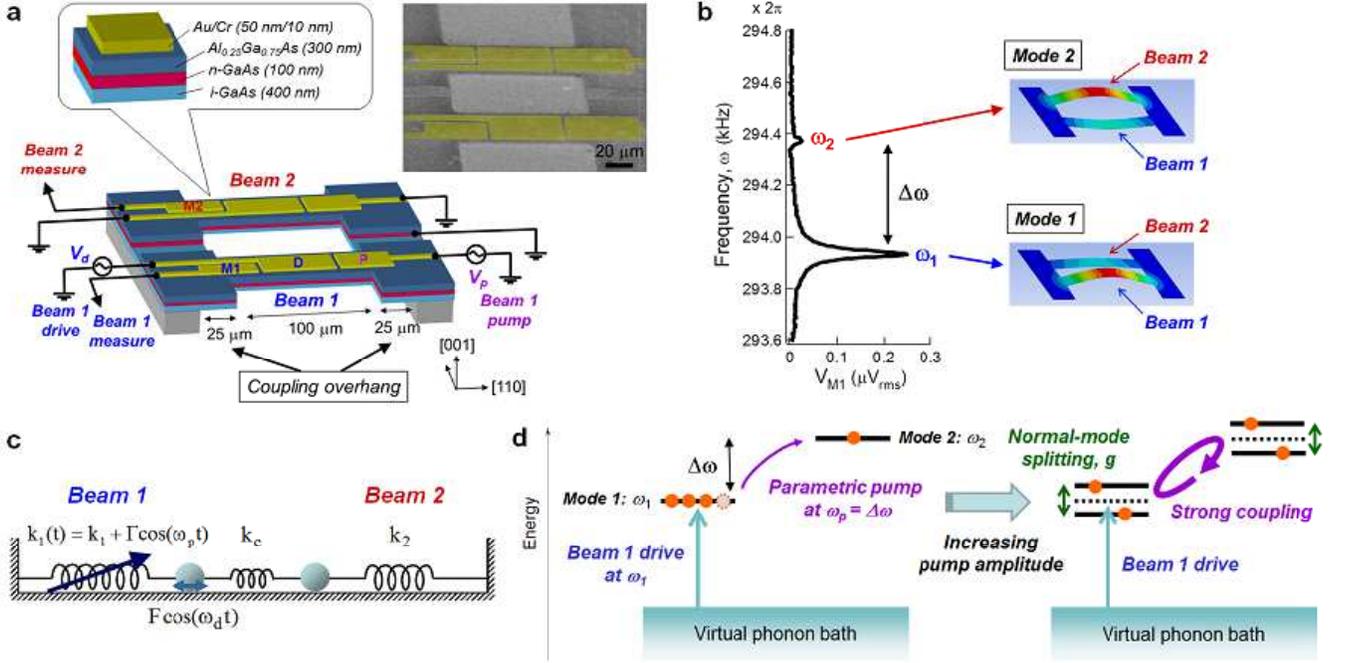}
\caption{{\bf Paired mechanical resonators and the parametric pumping scheme.} {\bf a,} Schematic drawing of the sample, the measurement and the false-colour scanning electron micrograph. The two doubly clamped beams are structurally interconnected via the coupling overhang. {\bf b,} Frequency response of beam 1 measured by driving it with $V_d = 1.5$ mV$_{\rm p-p}$ while the pump is deactivated ($V_p = 0$ V$_{\rm p-p}$). The mode shapes at $\omega_1$ and $\omega_2$ obtained by finite element method calculations are also shown (exaggerated for clarity). {\bf c,} Schematic drawing of the parametric pumping protocol in an equivalent spring model. $k_1$, $k_2$, and $k_c$ are respectively the spring constant of beam 1, beam 2, and the coupling overhang. $\Gamma$ represents the parametric pump amplitude. {\bf d,} Schematic drawing of the parametric pumping protocol in an energy diagram.}
\label{Fig1}
\end{center}
\end{figure*}

\begin{figure*}[tbp]
\begin{center}
\includegraphics[scale=0.59]{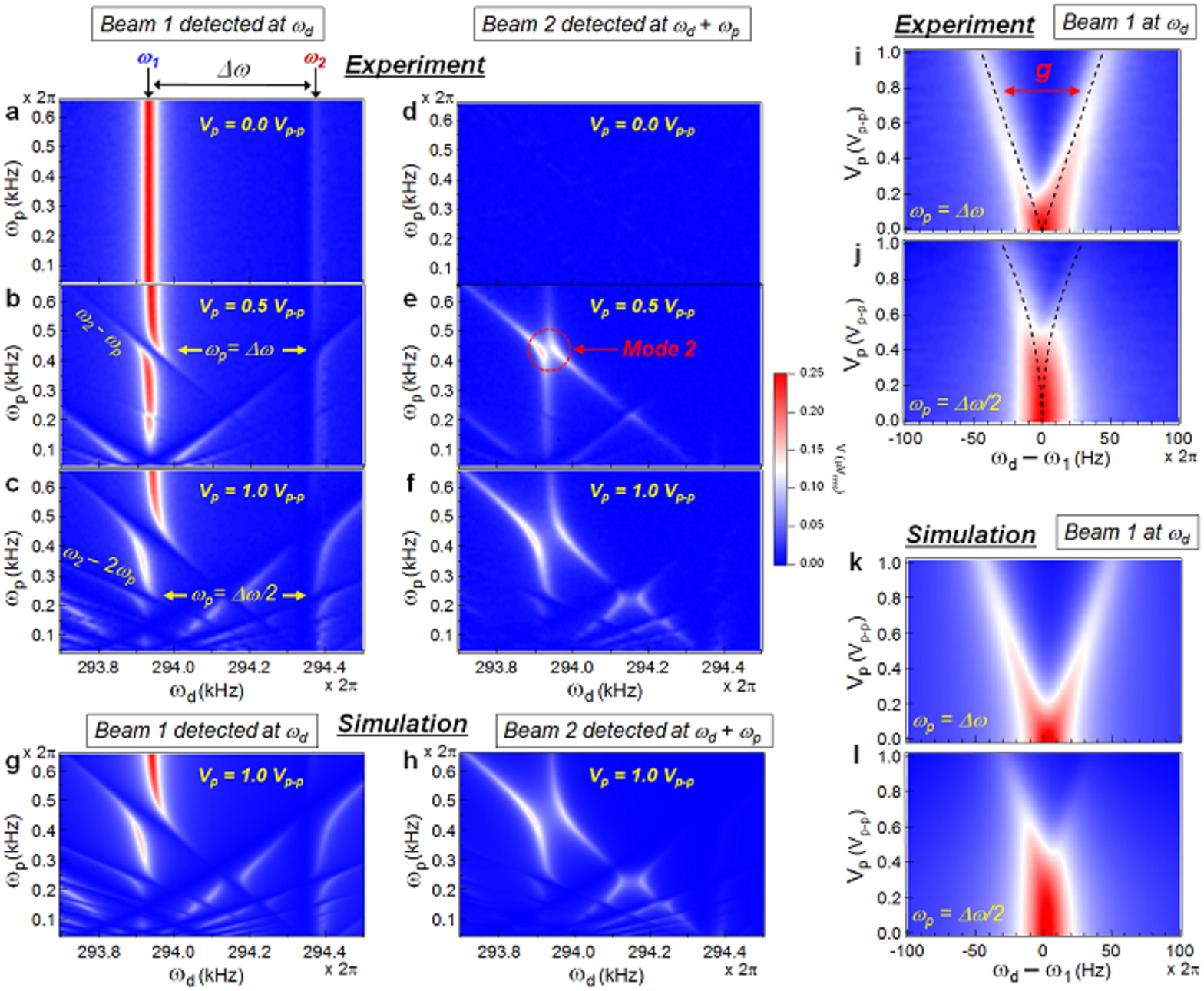}
\caption{{\bf Normal-mode splitting induced by parametric pumping.} {\bf a-c,} The drive frequency ($\omega_d$) and the pump frequency ($\omega_p$) response of beam 1 detected at frequency $\omega_d$ for three different pump voltages, $V_p$ = 0, 0.5 and 1.0 V$_{\rm p-p}$. {\bf d-f,} The $\omega_d$ and $\omega_p$ response of beam 2 detected at frequency $\omega_d + \omega_p$ for $V_p =$ 0, 0.5 and 1.0 V$_{\rm p-p}$. {\bf g,h,} Simulation results for ({\bf c}) and ({\bf f}), which were performed for the theoretical model expressed by Eqs. \ref{Eq1a} and \ref{Eq1b}. {\bf i,j,} The $V_p$ dependence of the splitting of mode 1 induced by the first- and second-order parametric coupling at $\omega_p = \Delta \omega$ and $\Delta \omega/2$, respectively. The broken curves in ({\bf i}) and ({\bf j}) represent the theoretical values of the mode splitting, which are proportional to $V_p$ and $(V_p)^2$ respectively (see Supplementary Information). The mode splitting corresponds to the coupling rate, $g$. {\bf k,l,} Simulation results for the splitting of mode 1 for $\omega_p = \Delta \omega$ and $\Delta \omega/2$, respectively.}
\label{Fig2}
\end{center}
\end{figure*}

The dynamic parametric coupling is developed in GaAs-based paired mechanical beams shown in Fig. \ref{Fig1}a, in which the piezoelectric effect is exploited to mediate all-electrical displacement transduction\cite{MahboobNPHYS}. The frequency response of beam 1 measured by harmonically driving it while the parametric pump is deactivated displays two coupled vibration modes (Fig. \ref{Fig1}b), where mode 1 ($\omega_1 = 2\pi \times 293.93$ kHz) is dominated by the vibration of beam 1 while mode 2 ($\omega_2 = 2\pi \times 294.37$ kHz) is dominated by the vibration of beam 2. The amplitude of mode 2 is much smaller than that of mode 1 reflecting the energy exchange due to the structural coupling via the overhang is inefficient because of the eigen-frequency difference between the two beams. This difference can be compensated by activating the parametric pump, which is induced by piezoelectrically modulating the spring constant of beam 1 with the pump frequency $\omega_p$ at around the frequency difference between the two modes, $\Delta\omega \equiv \omega_2 - \omega_1 $ (Fig. \ref{Fig1}c).

The dynamics of this system can then be expressed by the following equations of motion:
\begin{subequations}
\begin{eqnarray}
&& \ddot{x}_1 + \gamma_1\dot{x}_1 + [\omega_1^2 + \Gamma_1\cos(\omega_p t)] x_1 \nonumber \\
&& \;\;\;\;\;\;\;\;\;\;\;\;\;\;\;\;\; + \Lambda\cos(\omega_p t) x_2 = F_1\cos(\omega_d t + \phi) \label{Eq1a}\\
&& \ddot{x}_2 + \gamma_2\dot{x}_2 + [\omega_2^2 + \Gamma_2\cos(\omega_p t)] x_2 \nonumber \\
&& \;\;\;\;\;\;\;\;\;\;\;\;\;\;\;\;\;+ \Lambda\cos(\omega_p t) x_1 = F_2\cos(\omega_d t + \phi), \label{Eq1b}
\end{eqnarray}
\end{subequations}
where $x_i$ ($i = 1,2$) is the displacement of the $i$-th mode, $\omega_i$ is the mode frequency, $\gamma_i$ is the energy dissipation rate, $F_i$ is the drive force ($F_1 \gg F_2$), and $\omega_d$ is the drive frequency. When the frequency mismatch between mode 1 and mode 2 is compensated by activating the pump at $\omega_p \simeq \Delta\omega$, the terms containing $\Lambda$ transfer phonons (oscillations) from one mode to the other (Fig. \ref{Fig1}d). This {\it inter-modal} coupling can also be explained by the mixing of mode 1 (2) and the Stokes sideband, $\omega_2 - \omega_p$ (the anti-Stokes sideband, $\omega_1 + \omega_p$) of mode 2 (1) leading to the normal-mode splitting in the strong-coupling regime\cite{Weis,Groblacher,Safavi,TeufelCircuit,MahboobNPHYS} (Fig. \ref{Fig1}d). The equations also include terms proportional to $\Gamma_i$. These terms lead to {\it intra-modal} coupling which becomes significant for the higher-order parametric coupling as shown later. The above model can reproduce all the experimental results and is described in detail in Supplementary Information.

\begin{figure*}[tbp]
\begin{center}
\includegraphics[scale=0.5]{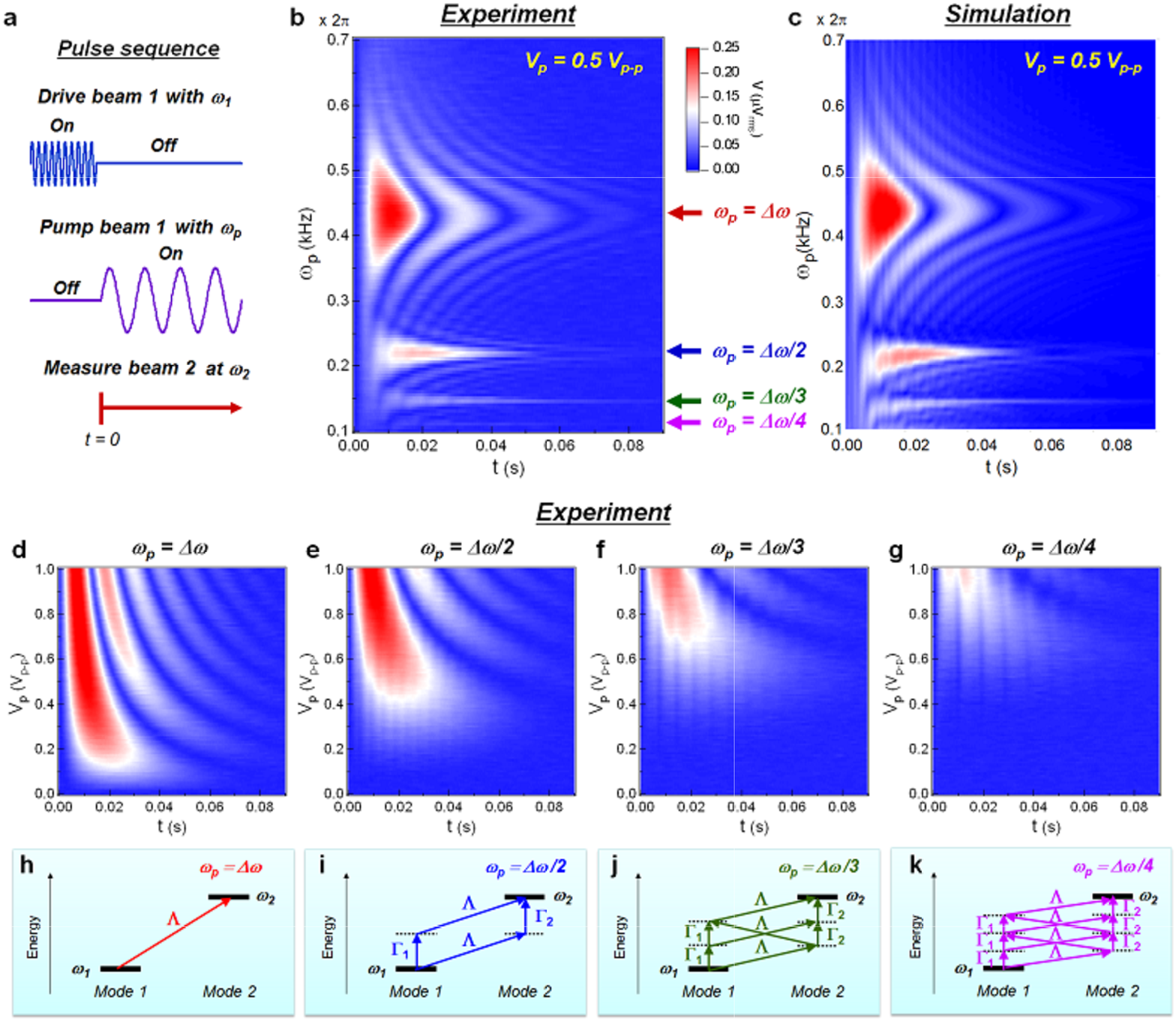}
\caption{{\bf Coherent energy exchange between the two beams/modes.} {\bf a,} Pulse sequence used for the measurements. {\bf b,c,} The $\omega_p$ dependence of the time-domain response of beam 2 at $\omega_2$ when the system is subjected to the above pulse sequence, from the experiment ({\bf b}) and the corresponding simulation ({\bf c}) for $V_p = 0.5$ V$_{\rm p-p}$. {\bf d-g,} The $V_p$ dependence of the time-domain response of beam 2 at $\omega_2$ for $\omega_p = \Delta \omega/n$, where $n =$ 1, 2, 3 and 4, respectively. The higher-frequency fringe patterns found for $n \geq 2$ are caused by the overlap of the coherent oscillations for the ($n$-1)-th order coupling, which is confirmed in ({\bf b}). {\bf h-k,} Schematic drawings of one, two, three, and four-pump phonon absorption processes, where $\Lambda$ is the inter-modal coupling coefficient while $\Gamma_1$ and $\Gamma_2$ are the intra-modal coupling coefficient for mode 1 and mode 2, respectively.}
\label{Fig3}
\end{center}
\end{figure*}

The parametric normal-mode splitting for modes 1 and 2 was experimentally confirmed by applying a pump with voltage $V_p$ at $\omega_p$ in addition to the weak harmonic driving at $\omega_d$ to beam 1 (Fig. \ref{Fig1}a). Probing the modes via beam 1 while the parametric pump is activated shows that mode 1 splits into two when $\omega_p \simeq \Delta \omega = 2\pi \times 0.44$ kHz clearly demonstrating the strong dynamical coupling (Figs. \ref{Fig2}a-\ref{Fig2}c). The phonon transfer from mode 1 to mode 2 can be confirmed in the response of beam 2 that was detected at frequency $\omega_d + \omega_p$ (Figs. \ref{Fig2}d-\ref{Fig2}f). The results indicate the emergence of its vibration and mode splitting at $\omega_d \sim \omega_1$ and $\omega_p \sim \Delta\omega$ showing that mode 2 ($\omega_2 = \omega_1 + \Delta\omega$) is excited by the dynamical coupling. A phonon reaction picture can help to understand this elementary process where phonons are created in mode 2 at the expense of probe phonons in mode 1 and pump phonons, i.e., via the one-pump phonon absorption process, $\hbar\omega_1 + \hbar\omega_p \rightarrow \hbar\omega_2$ (Figs. \ref{Fig1}d and \ref{Fig3}h). Together with the reversed emission process, $ \hbar\omega_2 \rightarrow \hbar\omega_1 + \hbar\omega_p$, strong coupling is established (Fig. \ref{Fig1}d).

The $V_p$ dependence of the mode splitting at $\omega_p = \Delta \omega$ shows that the coupling strength is highly controllable with the splitting being proportional to $V_p$ (Fig. \ref{Fig2}i). This linear $V_p$ dependence is due to the fact that the inter-modal coupling coefficient, $\Lambda$, is proportional to $V_p$, which can be theoretically reproduced by Eqs. \ref{Eq1a} and \ref{Eq1b} (Fig. \ref{Fig2}k). The separation between the split peaks provides the coupling rate, $g$, which can become strong enough to exceed the intrinsic energy dissipation rate of the two modes ($\gamma_1 \simeq \gamma_2 = 2\pi \times 22$ Hz) by more than a factor of four ($g = 2\pi \times 90$ Hz for $V_p = 1.0$ $V_{\rm p-p}$), i.e., ultra-strong parametric coupling can be achieved in this system.

More remarkably additional mode splitting, in which the pump frequency does not correspond to the frequency difference between the two modes, can also be observed. The additional splitting occurs when $\omega_p \simeq \Delta \omega/2 = 2\pi \times 0.22$ kHz for both modes 1 and 2 (Fig. \ref{Fig2}c). This splitting is caused by a second-order parametric coupling via two-pump phonon absorption/emission process, i.e., $\hbar\omega_1 + 2\hbar\omega_p \leftrightarrow \hbar\omega_2$, which leads to dynamic coupling between mode 1 (2) and the second Stokes sideband, $\omega_2 - 2\omega_p$ (the second anti-Stokes sideband, $\omega_1 + 2\omega_p$) of mode 2 (1). The $V_p$ dependence of this mode splitting indicates that it has a parabolic dependence (Fig. \ref{Fig2}j). This is because the second-order process requires a two-step phonon excitation path from mode 1 to mode 2, and vice versa, through the intermediary energy level $(\omega_1 + \omega_2)/2$ via both the intra-modal coupling ($\Gamma_i \propto V_p$) and the inter-modal coupling ($\Lambda \propto V_p$), i.e., $[\Gamma_i \times \Lambda] \propto V_p^2$, as shown in Fig. \ref{Fig3}i. Again, the corresponding mode splitting shows good agreement with the theoretical simulations (Fig. \ref{Fig2}l). Uniquely, the second-order phonon processes can be observed in the present system owing to the strong parametric coupling between the two resonators/modes.

\begin{figure}[tbp]
\begin{center}
\includegraphics[scale=0.55]{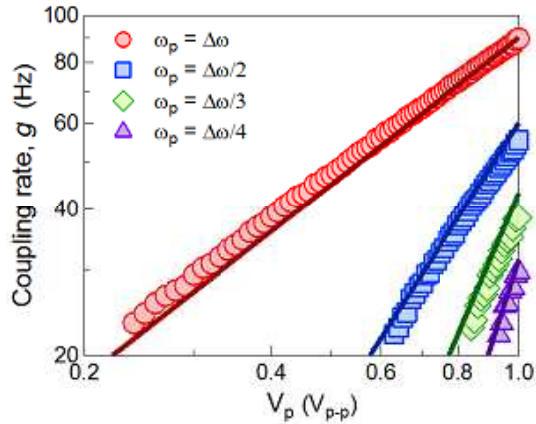}
\caption{{\bf Coupling rate for the $n$-th order parametric pumping.} The logarithmic plots of the coupling rate, $g$, for $\omega_p = \Delta \omega/n$ ($n =$ 1, 2, 3 and 4) with respect to $V_p$, obtained from the Fourier transforms of the time-domain response shown in Figs. \ref{Fig3}d-\ref{Fig3}g. Solid lines represent the theoretical values, which are proportional to $(V_p)^{n}$ (see Supplementary Information).}
\label{Fig4}
\end{center}
\end{figure}

The ultra-strong parametric coupling in the paired mechanical resonators opens up a path to coherent control of the mechanical oscillations. The time-domain measurements utilizing the pulse sequence shown in Fig. \ref{Fig3}a and described in Methods Summary enable us to observe coherent and periodic energy exchange between the two beams/modes. The pump frequency dependence of the time-domain response of beam 2 at $\omega_2$ clearly shows the periodic amplitude oscillations at $\omega_p \simeq \Delta\omega$ for the first-order ($n = 1$) parametric coupling (Fig. \ref{Fig3}b). The $V_p$ dependence at $\omega_p = \Delta\omega$ shows that the vibration energy of mode 1 (beam 1) can be transferred to mode 2 (beam 2) and back eight times at $V_p = 1.0$ V$_{\rm p-p}$ (Fig. \ref{Fig3}d). Coherent energy exchange for the second-order ($n = 2$) parametric coupling can also be observed at $\omega_p \simeq \Delta \omega/2$ (Fig. \ref{Fig3}b), where up to five oscillation periods are observed in the range of $V_p \leq$ 1.0 V$_{\rm p-p}$ (Fig. \ref{Fig3}e). The coupling rate $g$, extracted from the Fourier transforms of the time-domain response is proportional to $V_p$ for the first-order coupling and $(V_p)^2$ for the second-order coupling (Fig. \ref{Fig4}). These coupling rates correspond perfectly to the mode splitting observed in the frequency response measurement. 

Even more surprisingly, the time-domain measurements enable us to observe higher-order parametric coupling, i.e., $n \geq 3$. Figures \ref{Fig3}f and \ref{Fig3}g show the coherent energy exchange between the two beams/modes for $\omega_p = \Delta \omega/3 = 2\pi \times 0.147$ kHz and $\omega_p = \Delta \omega/4 = 2\pi \times 0.11$ kHz, respectively. These coherent oscillations are caused by the $n$-pump phonon absorption/emission processes, i.e. $\hbar\omega_1 + n\hbar\omega_p \leftrightarrow \hbar\omega_2$, through the intermediary energy levels via intra- and inter-modal coupling, e.g., $[\Gamma_1 \times \Gamma_1 \times \Lambda]$, as shown in Figs. \ref{Fig3}j and \ref{Fig3}k. The Fourier transforms of the time-domain response reveal that the coupling rate exhibits $(V_p)^{n}$ dependence even for $n \geq 3$ (Fig. \ref{Fig4}), which again shows good agreement with the theoretical model. 

The present results show that electromagnetic pulse techniques, which are commonly used to coherently manipulate quantum two-level systems\cite{O'Connell,Martinis}, can also be applied to coherently control mechanical systems, which enables the realization of superposition states between the two coupled vibration modes. By tuning the parametric-pump frequency, multi-wave phonon mixing involving an arbitrary number of pump phonons can be achieved in an analogous fashion to multi-wave photon mixing\cite{KippenbergScience}. The parametric pumping allows highly controllable time-domain manipulation of the phonon population with the adjustment of the pump-pulse duration, thus permitting $\pi$ and $\pi/2$-pulse operations on the Bloch sphere\cite{O'Connell} (see also independent experiments at LMU with a single mechanical resonator\cite{Weig}). This coherent control further expands the applications of mechanical resonators, including the high-speed operation of high-$Q$ mechanical resonators\cite{Yamaguchi} and mechanical logic operations\cite{MahboobNNano}. Although the system demonstrated here is in the classical regime with large mode occupation, these techniques can also be extended to the quantum regime\cite{O'Connell,KippenbergNATURE,TeufelGround}, leading to the exciting possibility of quantum-coherent coupling and entanglement between two distinct macroscopic mechanical objects\cite{Wineland,Harlander,Martinis}. 

\section*{METHODS SUMMARY}
{\bf Fabrication.} The sample was fabricated by photolithography from a heterostructure consisting of Al$_{0.25}$Ga$_{0.75}$As, Si-doped $n$-GaAs, undoped $i$-GaAs and 2-$\mu$m-thick Al$_{0.65}$Ga$_{0.35}$As sacrificial layers grown on a GaAs(001) substrate by molecular beam epitaxy (Fig. \ref{Fig1}a). AuGeNi was deposited on the supporting part to obtain an ohmic contact to the conductive $n$-GaAs layer, while Au/Cr gates were formed on the top of the beams (Fig. \ref{Fig1}a). The suspended structure was completed by deep mesa and isotropic sacrificial layer etching, where the 40-$\mu$m-separated beams were electrically isolated by the shallow mesa etch. All the measurements were done by setting the sample in a vacuum (5 $\times$ 10$^{-5}$ Pa) at 1.5 K with a $^{\rm 4}$He cryostat. 

{\bf Frequency-domain measurement.} Beam 1 was driven by a continuous AC voltage ($V_d$) applied to gate D with the frequency $\omega_d \sim \omega_1$. Parametric pumping was achieved by applying a continuous AC pump voltage ($V_p$) to gate P with the frequency $\omega_p \sim \omega_2 - \omega_1$. The responses of beam 1 and beam 2 were simultaneously monitored via gates M1 and M2, where the signal was amplified with a room-temperature low-noise pre-amplifier (NF:SA-220F5) and detected with a lock-in amplifier (SRS:SR844) (see Supplementary Information).

{\bf Time-domain measurement.} A sinusoidal drive pulse voltage ($V_d$) with $\omega_d = \omega_1$ was applied to gate D with the pulse period of 0.1 s. A sinusoidal pump pulse with $\omega_p = (\omega_2 - \omega_1)/n$ was applied to gate P at the moment the drive pulse was switched off, where the pump pulse also has the pulse period of 0.1 s. The time response of beam 2 at $\omega_2$ was measured through gate M2 with an oscilloscope (Agilent:DSO90254A) via the pre-amplifier and the lock-in detector (see Supplementary Information). The data was averaged 20 times with the oscilloscope. 

\section*{Acknowledgements} H.O. thanks A. Taspinar for supporting the data analysis. The authors acknowledge T. Faust, J. P. Kotthaus and E. M. Weig for their critical reading of the manuscript. This work was partly supported by JSPS KAKENHI (23241046 \& 20246064).\\*

\setcounter{section}{0} %set all counters to 0, the first item is then Fig. 1 etc...
\setcounter{figure}{0}
\setcounter{equation}{0}
\onecolumngrid %switch to single-column layout \clearpage %leave the rest of the page blank and start a new one

\newpage
\renewcommand{\thefigure}{S\arabic{figure}}
\renewcommand{\theequation}{S\arabic{equation}}

\section*{\large Supplementary Information for\\
"Coherent phonon manipulation in coupled mechanical resonators"}

\begin{figure}[bp]
\begin{center}
\includegraphics[scale=0.8]{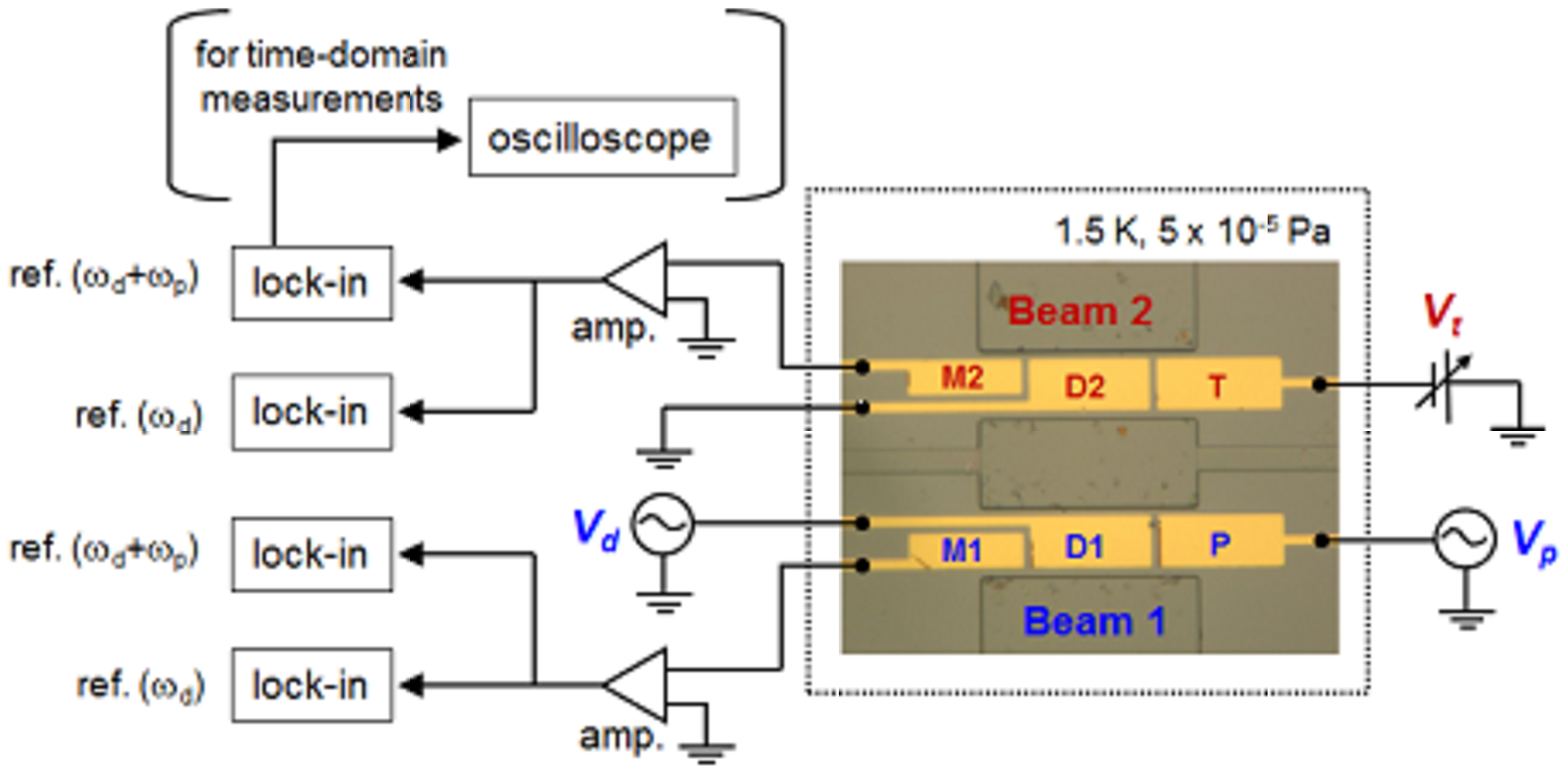}
\caption{A topographic optical image of the paired mechanical resonators and a schematic of the circuit for the measurements.}
\label{FigS1}
\end{center}
\end{figure}

\section{Measurement scheme}
Piezoelectrically frequency tunable paired mechanical resonators were prepared to demonstrate parametric coupling (Fig. \ref{FigS1}). Beam 1 and beam 2 are structurally (elastically) coupled via overhang parts (see Fig. 1a in the main text) that were formed by the sacrificial layer etching process. The structural-coupling constant can be estimated from the results of the DC detuning experiments (described in the next section), where gate T is used to change the eigenfrequency of beam 2 with the application of DC voltage ($V_t$) using a function generator (NF:WF1974) (Fig. \ref{FigS1}). For the parametric pumping measurements, both the drive voltage and pump voltage were applied to beam 1 with the function generator. A weak AC drive voltage ($V_d = 1.5$ mV$_{\rm p-p}$) was applied to gate D1 while the conductive $n$-GaAs layer (see Fig. 1a in the main text) was grounded via the ohmic contact formed on the supporting part (not shown). An AC pump voltage ($V_p$) was applied to gate P, where the pump frequency ($\omega_p$) was set at around the frequency difference between the two coupled vibration modes ($\omega_2 - \omega_1$), while $V_t$ was set to 0 V. In the frequency-domain measurements, the responses of beam 1 and beam 2 were simultaneously monitored via strain-induced piezoelectric voltage at gate M1 and gate M2 with low-noise pre-amplifiers (NF:SA-220F5) and lock-in amplifiers (SRS:SR844). Here, we used two lock-in amplifiers to measure each beam's response, one to measure the signal at the drive frequency $\omega_d$ and the other to measure the idler at the frequency $\omega_d + \omega_p$ as shown in Fig. \ref{FigS1}. In contrast, in the time-domain measurements, the real-time response of beam 2 at the idler frequency $\omega_d + \omega_p$ was monitored with an oscilloscope (Agilent:DSO90254A) via a lock-in amplifier as shown in Fig. \ref{FigS1}. 

\begin{figure}[tbp]
\begin{center}
\includegraphics[scale=0.7]{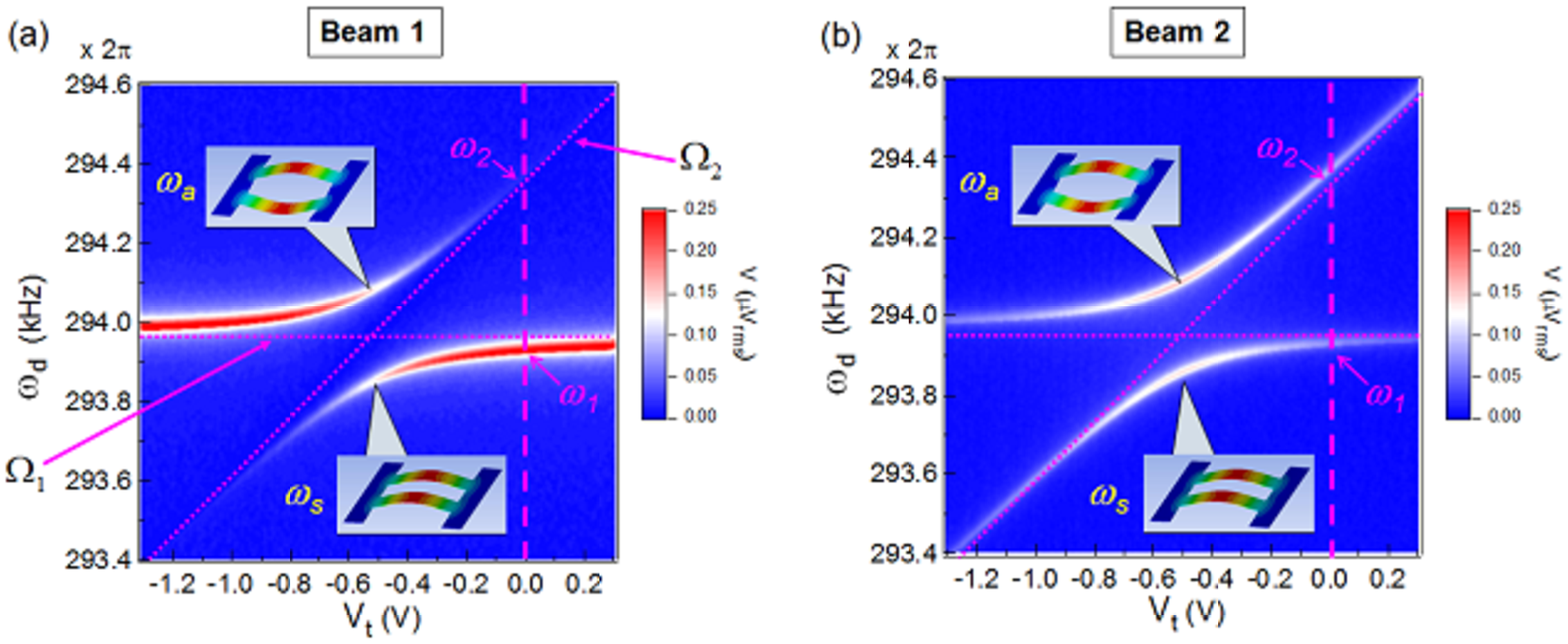}
\caption{DC detuning voltage ($V_t$) dependence of the frequency response of (a) beam 1 and (b) beam 2 measured by harmonically driving beam 1 with no parametric pumping. Dotted lines show the eigenfrequency of beam 1 ($\Omega_1$) and beam 2 ($\Omega_2$). The response at $V_t = 0$ V is highlighted by a broken line. Drawings of the symmetric vibration and anti-symmetric vibration for the perfect tuning point are also shown.}
\label{FigS2}
\end{center}
\end{figure}

\section{Structural coupling}

In the paired mechanical resonators, the degree of vibration coupling strongly depends on the eigenfrequency difference between the two resonators. This was confirmed by applying DC detuning voltage ($V_t$) to beam 2 while beam 1 was harmonically driven (Fig. \ref{FigS1}). The $V_t$ dependence of the frequency response of beam 1 and beam 2 for $V_p =$ 0 V$_{\rm p-p}$ is shown in Figs. \ref{FigS2}a and \ref{FigS2}b, respectively. At $V_t = 0$ V, the amplitude of mode 1 ($\omega_1 = 2\pi \times 293.93$ kHz) is much larger than that of mode 2 ($\omega_2 = 2\pi \times 294.37$ kHz) in beam 1 (Fig. \ref{FigS2}a). This is because the eigenfrequency of beam 2 ($\Omega_2$) does not coincide with that of beam 1 ($\Omega_1$) at $V_t = 0$ V. The separation of the two mode frequencies reduces with the application of the negative $V_t$ and it is minimized at $V_t = -0.5$ V, showing the avoided crossing of the two modes (Figs. \ref{FigS2}a and \ref{FigS2}b). This is because $\Omega_2$, which is initially higher than $\Omega_1$ owing to the fabrication error, decreases with increasing negative $V_t$ due to the piezoelectric effect and perfect tuning ($\Omega_2 = \Omega_1$) is achieved at $V_t = -0.5$ V. In the perfectly tuned condition, the amplitudes of the two modes (and the two beams) coincide with each other, where the lower-frequency mode ($\omega_s = 2\pi \times 293.85$ kHz) corresponds to the symmetric vibration and the higher-frequency mode ($\omega_a = 2\pi \times 294.08$ kHz) corresponds to the anti-symmetric vibration. The difference between the two mode frequencies at the perfect tuning point provides information about the structural coupling constant ($c = \Omega_c^2$), where $\Omega_c = \sqrt{(\omega_a^2-\omega_s^2)/2} = 2\pi \times 8.22$ kHz is 2.8 $\%$ of $\Omega_1$. From Fig. \ref{FigS2}a (and \ref{FigS2}b) the piezoelectric frequency detuning coefficient is estimated to be $\delta\Omega_2/\delta V_t = 2\pi \times$ 0.75 kHz/V.

\section{Theoretical simulations}

\subsection{Model}

We used a standard model of two coupled harmonic oscillators with the linearly voltage-controlled detuning of resonance frequency. The equations of motion are given by
\begin{subequations}
\begin{eqnarray}
& & \ddot{X}_1 + \gamma_1\dot{X}_1 + [\Omega_1^2 + \alpha V_1(t)]X_1 = c(X_2 - X_1) + F_1(t) \label{EqS1a}\\
& & \ddot{X}_2 + \gamma_2\dot{X}_2 + [\Omega_2^2 + \alpha V_2(t)]X_2 = c(X_1 - X_2) + F_2(t) \label{EqS1b}. 
\end{eqnarray}
\end{subequations}
Here, $X_i$ ($i = 1,2$) is the displacement of the $i$-th oscillator, $\Omega_1$ and $\Omega_2 = \Omega_1 + \delta\Omega$ ($\Omega_1 >> \delta\Omega$) are the resonance frequencies of beam 1 and beam 2, respectively, $\gamma_i = \Omega_i/Q_i$ is the energy dissipation (damping) rate where $Q_i$ is the quality factor, $V_i$ is the detuning gate voltage, $F_i$ is the harmonic driving force, and $c$ is the geometrical coupling constant between the two oscillators. The term $\alpha V_i(t)$ gives the eigenfrequency detuning and the device we used has the detuning coefficient of $\alpha/2\Omega_i = \delta \Omega_i/\delta V = 2\pi \times$ 0.75 kHz/V (Figs. \ref{FigS2}a and \ref{FigS2}b). We describe the case when beam 1 is harmonically driven at frequency $\omega_d$ with amplitude $F_1 = F$, the eigenfrequency of beam 1 ($\Omega_1$) is parametrically modulated (pumped) at frequency $\omega_p$ with a voltage amplitude of $V_p = \Gamma/\alpha$, and no detuning gate voltage is applied to beam 2. Then the equations of motion become
\begin{subequations}
\begin{eqnarray}
& & \ddot{X}_1 + \gamma\dot{X}_1 + [\Omega_1^2 + \Gamma\cos(\omega_p t)]X_1 = c(X_2 - X_1) + F\cos(\omega_d t + \phi) \label{EqS2a}\\
& & \ddot{X}_2 + \gamma\dot{X}_2 + [\Omega_1^2 + 2\Omega_1\delta\Omega]X_2 = c(X_1 - X_2) \label{EqS2b}, 
\end{eqnarray}
\end{subequations}
where $\phi$ is the phase factor determined from the phase relation between the harmonic driving $F\cos(\omega_d t)$ and the parametric pump $\Gamma\cos(\omega_p t)$, and the damping rate was assumed to be $\gamma = \gamma_1 = \gamma_2$. When no parametric pump is applied, the equations become
\begin{subequations}
\begin{eqnarray}
& & \ddot{X}_1 + \gamma\dot{X}_1 + (\Omega_1^2 + c)X_1 -cX_2 = F\cos(\omega_d t + \phi) \label{EqS3a}\\
& & \ddot{X}_2 + \gamma\dot{X}_2 + (\Omega_1^2 + 2\Omega_1\delta\Omega + c)X_2 -cX_1 = 0 \label{EqS3b}, 
\end{eqnarray}
\end{subequations}
or
\begin{eqnarray}
\left( \frac{d^2}{dt^2} + \gamma\frac{d}{dt} + \Omega_1^2 + c + \Omega_1\delta\Omega \right) 
\left( 
\begin{array}{cc}
X_1 \\
X_2 \\
\end{array}
\right) 
-
\left(
\begin{array}{cc} 
\Omega_1\delta\Omega & c \\
c & -\Omega_1\delta\Omega \\
\end{array}
\right) 
\left( 
\begin{array}{cc}
X_1 \\
X_2 \\
\end{array}
\right) \nonumber \\
\;\;\;\;\;\;\;\;\;\;\;\;\;\;\;\;\;\;\;\;\;\;\;\;\;\;\;\;\;\;\;\;\;\;\;\;\;\;\;\;\;\;\;\;\;\;\;\;\;\;\;\;\;\;\;\;\;\;\;\;\;\;\;\;\;\;\;\;
= \left( 
\begin{array}{cc} 
F\cos(\omega_dt + \phi) \\
0 \\
\end{array}
\right). \label{EqS4}
\end{eqnarray}
Using the diagonalization orthogonal matrix {\bf U} defined by
\begin{eqnarray}
{\bf U} = \frac{1}{\sqrt{2\lambda}}
\left( 
\begin{array}{cc}
\sqrt{\lambda + d} & \sqrt{\lambda - d} \\
\sqrt{\lambda - d} & -\sqrt{\lambda + d} \\
\end{array}
\right), \label{EqS5}
\end{eqnarray}
where $d = \Omega_1\delta\Omega$ and $\lambda = \sqrt{c^2 + \Omega_1^2\delta\Omega^2}$. Eq. \ref{EqS5} is diagonalized as
\begin{eqnarray}
\left( \frac{d^2}{dt^2} + \gamma\frac{d}{dt} + \Omega_1^2 + c + \Omega_1\delta\Omega \right) 
\left( 
\begin{array}{cc}
x_1 \\
x_2 \\
\end{array}
\right) 
-
\left(
\begin{array}{cc} 
\lambda & 0 \\
0 & -\lambda \\
\end{array}
\right) 
\left( 
\begin{array}{cc}
x_1 \\
x_2 \\
\end{array}
\right)
= 
\left( 
\begin{array}{cc} 
U_{11} \\
U_{21} \\
\end{array}
\right)
F\cos(\omega_dt + \phi), \label{EqS6}
\end{eqnarray}
where $x_i$ is the displacement of the $i$-th mode given by
\begin{eqnarray}
\left(
\begin{array}{cc}
x_1 \\
x_2 \\
\end{array}
\right)
= {\bf U}
\left(
\begin{array}{cc}
X_1 \\
X_2 \\
\end{array}
\right). \label{EqS7}
\end{eqnarray}
The equivalent expression of Eq. \ref{EqS6} is
\begin{subequations}
\begin{eqnarray}
& & \ddot{x}_1 + \gamma\dot{x}_1 + \omega_1^2x_1 = U_{11}F\cos(\omega_d t + \phi), \;\;\;\;\; \omega_1^2 = \Omega_1^2 + c + d - \sqrt{c^2 + d^2} \label{EqS8a}\\
& & \ddot{x}_2 + \gamma\dot{x}_2 + \omega_2^2x_2 = U_{21}F\cos(\omega_d t + \phi), \;\;\;\;\; \omega_2^2 = \Omega_1^2 + c + d + \sqrt{c^2 + d^2} \label{EqS8b}. 
\end{eqnarray}
\end{subequations}
The two independent modes are referred as mode 1 and mode 2 hereafter. 

In the case when the parametric pumping is ON, Eqs. \ref{EqS2a} and \ref{EqS2b} can be expressed by using the two modes as
\begin{subequations}
\begin{eqnarray}
& & \ddot{x}_1 + \gamma\dot{x}_1 + \omega_1^2x_1 + (\Gamma_1 x_1 + \Lambda x_2)\cos(\omega_p t) = U_{11}F\cos(\omega_d t + \phi) \label{EqS9a}\\
& & \ddot{x}_2 + \gamma\dot{x}_2 + \omega_2^2x_2 + (\Gamma_2 x_2 + \Lambda x_1)\cos(\omega_p t) = U_{21}F\cos(\omega_d t + \phi) \label{EqS9b}, 
\end{eqnarray}
\end{subequations}
where $\Gamma_1$, $\Gamma_2$, and $\Lambda$ are given by
\begin{eqnarray}
& & \Gamma_1 = \Gamma U_{11}^2 = \frac{\Gamma}{2}\left( 1 + \frac{\Omega_1\delta\Omega}{\sqrt{c^2 + \Omega_1^2\delta\Omega^2}} \right) \nonumber \\
& & \Gamma_2 = \Gamma U_{21}^2 = \frac{\Gamma}{2}\left( 1 - \frac{\Omega_1\delta\Omega}{\sqrt{c^2 + \Omega_1^2\delta\Omega^2}} \right) \nonumber \\
& & \Lambda = \Gamma U_{11}U_{21} = \frac{\Gamma c}{2\sqrt{c^2 + \Omega_1^2\delta\Omega^2}}. \nonumber
\end{eqnarray}
By simplifying Eqs. \ref{EqS9a} and \ref{EqS9b} with $F_1 = U_{11}F$ and $F_2 = U_{21}F$, we obtain the forms of Eq. 1 in the main text,
\begin{subequations}
\begin{eqnarray}
&& \ddot{x}_1 + \gamma\dot{x}_1 + [\omega_1^2 + \Gamma_1\cos(\omega_p t)] x_1  + \Lambda\cos(\omega_p t) x_2 = F_1\cos(\omega_d t + \phi) \;\;\;\;\;\;\;\;\;\;\;\;\;\;\;\;\;\;{\rm (1a)}\nonumber\\
&& \ddot{x}_2 + \gamma\dot{x}_2 + [\omega_2^2 + \Gamma_2\cos(\omega_p t)] x_2  + \Lambda\cos(\omega_p t) x_1 = F_2\cos(\omega_d t + \phi). \;\;\;\;\;\;\;\;\;\;\;\;\;\;\;\;\;{\rm (1b)}\nonumber
\end{eqnarray}
\end{subequations}
The terms including $\Lambda$ represent the {\it inter-modal} coupling, which transfers phonons from one mode to the other. On the other hand, the terms containing $\Gamma_i$ represent the {\it intra-modal} coupling, which becomes significant for the higher-order parametric coupling as described in the latter section. Because $\Gamma = \alpha V_p$, both the intra-modal coupling coefficient $\Gamma_i$ and the inter-modal coupling coefficient $\Lambda$ are proportional to the pump voltage $V_p$. Note that $\Lambda$ is maximized when $\delta\Omega = 0$ but still sufficiently large when $\delta\Omega$ is comparable to $c/\Omega_1$, where $c \simeq 2.6 \times 10^9$ Hz$^2$, $\Omega_1 \simeq 1.8 \times 10^6$ Hz, and $\delta\Omega = 2.5 \times 10^3$ Hz in the present paired mechanical resonators; therefore, $\delta\Omega \sim c/\Omega_1$. 

\begin{figure}[tbp]
\begin{center}
\includegraphics[scale=0.98]{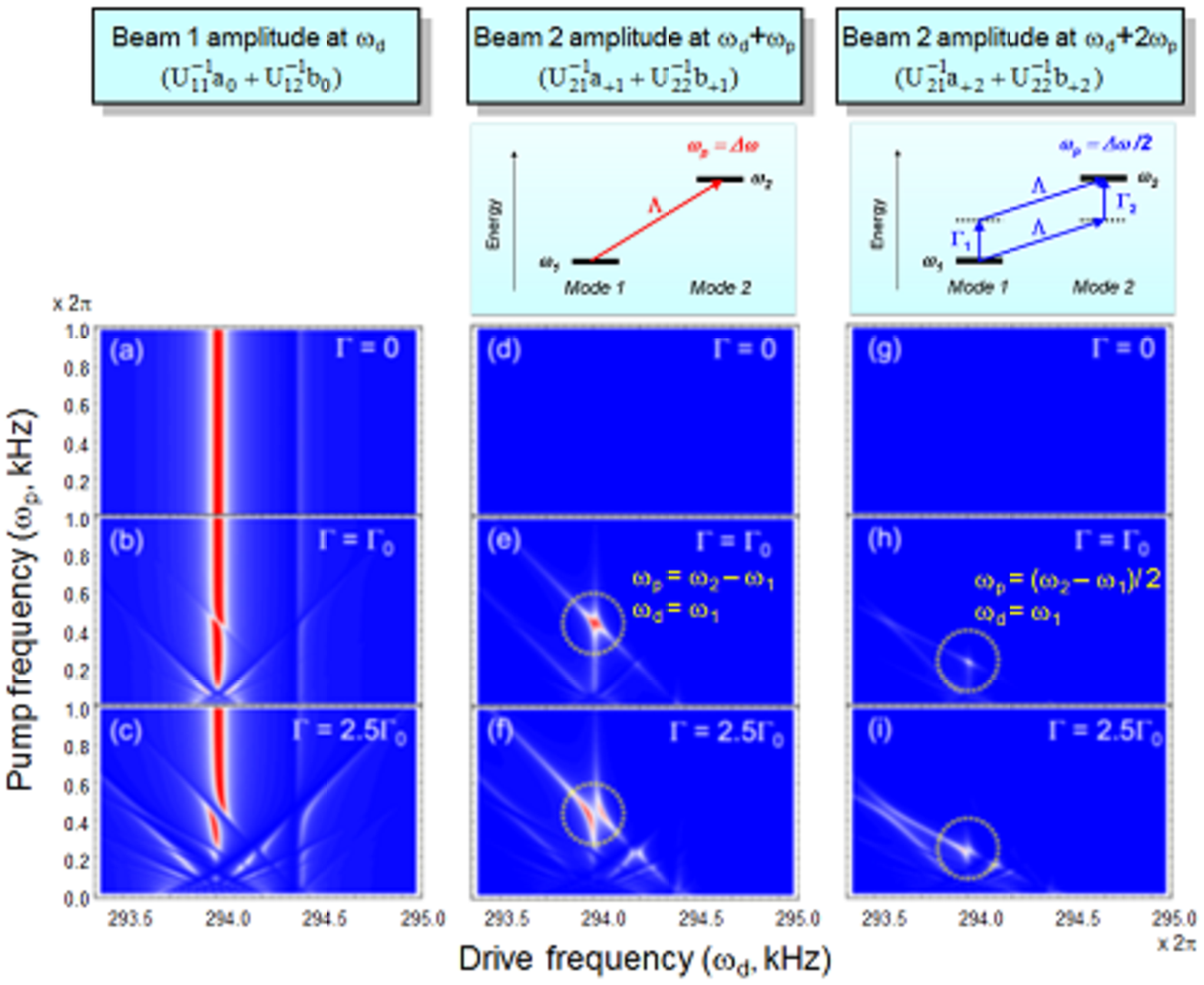}
\caption{Numerical simulations of oscillation amplitudes of [(a)-(c)] beam-1 at drive frequency $\omega_d$, [(d)-(f)] beam-2 at the first-order idler frequency $\omega_d + \omega_p$, and [(g)-(i)] beam-2 at the second-order idler frequency $\omega_d + 2\omega_p$ as a function of pump and drive frequency.}
\label{FigS3}
\end{center}
\end{figure}

\subsection{Higher-idler rotating-frame approximation}

Equations 1a and 1b can be numerically solved by using rotating-frame approximation with higher-harmonic idler components. By introducing slowly varying $m$-th harmonic idler amplitudes, $a_m(t)$ and $b_m(t)$ with $m$ as any integer, $x_1(t)$ and $x_2(t)$ can be expanded as
\begin{subequations}
\begin{eqnarray}
& & x_1(t) = Re\left[ \sum^{+\infty}_{m = -\infty}a_m(t)e^{i(\omega + m\omega_p)t}\right] \label{EqS10a} \\
& & x_2(t) = Re\left[ \sum^{+\infty}_{m = -\infty}b_m(t)e^{i(\omega + m\omega_p)t}\right]. \label{EqS10b}
\end{eqnarray}
\end{subequations}
Then, the equations of motion become a series of equations:
\begin{subequations}
\begin{eqnarray}
\left[ 2i(\omega_d + m\omega_p) + \gamma\right]\dot{a}_m &+& \left[ -(\omega_d + m\omega_p)^2 + i\gamma(\omega_d + m\omega_p) + \omega_1^2\right]a_m \nonumber \\
&+& \frac{\Gamma_1}{2}a_{m+1} + \frac{\Gamma_1}{2}a_{m-1} + \frac{\Lambda}{2}b_{m+1} + \frac{\Lambda}{2}b_{m-1} = Fe^{i\delta}\delta_{m,0} \label{EqS11a} \\
\left[ 2i(\omega_d + m\omega_p) + \gamma\right]\dot{b}_m &+& \left[ -(\omega_d + m\omega_p)^2 + i\gamma(\omega_d + m\omega_p) + \omega_2^2\right]b_m \nonumber \\
&+& \frac{\Gamma_2}{2}b_{m+1} + \frac{\Gamma_2}{2}b_{m-1} + \frac{\Lambda}{2}a_{m+1} + \frac{\Lambda}{2}a_{m-1} = 0, \label{EqS11b} 
\end{eqnarray}
\end{subequations}
where the second-order time derivatives of $a_m$ and $b_m$ were neglected. The solution in $N$-th order approximation can be obtained by solving the equations by neglecting the high-order idler, $a_m$ and $b_m$ for $m > N$.

\subsection{Simulations of frequency response and idler spectrum}

The frequency response and the idler spectrum, i.e., the time-independent solutions, can be obtained by neglecting the first terms on the left-hand side of Eqs. \ref{EqS11a} and \ref{EqS11b}. The solution for $4N+2$ vibration amplitudes ${a_m, b_m}$ can be obtained by calculating the inverse of the $(4N+2)\times(4N+2)$ matrix. The details of the calculation will be reported elsewhere and we here only show the results of ninth-order approximation ($N=9$) for the frequency response of beam 1 ($U_{11}^{-1}a_0 + U_{12}^{-1}b_0$) and the first- and second-order idler amplitude of beam 2 ($U_{21}^{-1}a_{+1} + U_{22}^{-1}b_{+1}$ and $U_{21}^{-1}a_{+2} + U_{22}^{-1}b_{+2}$) with three different pump intensities in Figs. \ref{FigS3}a-\ref{FigS3}i. Here, we introduced the unit pump intensity $\Gamma_0 = 2\pi\omega_1^2/Q$. The condition of strong coupling is given by $\Gamma \geq \Gamma_0$. The avoided crossing was observed in the frequency response when $\omega_p = \omega_2 - \omega_1$ (Fig. \ref{FigS3}c). This coupling results in phonons being created in mode 2 ($\omega_2$) at the expense of probe phonons in mode 1 ($\omega_1$) and the pump phonons ($\omega_p$), i.e., via the one-pump phonon absorption process, $\hbar\omega_1 + \hbar\omega_p \rightarrow \hbar\omega_2$. The creation of phonons at $\omega_2$ is confirmed by the first-order idler amplitude, which appears when $\omega_d = \omega_1$ and $\omega_p = \omega_2 - \omega_1$ (Figs. \ref{FigS3}e and \ref{FigS3}f). On the other hand, the second-order parametric coupling via the two-pump phonon absorption process, $\hbar\omega_1 + 2\hbar\omega_p \rightarrow \hbar\omega_2$, appears when $\omega_p = (\omega_2 - \omega_1)/2$. This process is confirmed by the second-order idler amplitude, which appears when $\omega_d = \omega_1$ and $\omega_p = (\omega_2 - \omega_1)/2$ (Figs. \ref{FigS3}h and \ref{FigS3}i).

\begin{figure}[tbp]
\begin{center}
\includegraphics[scale=0.59]{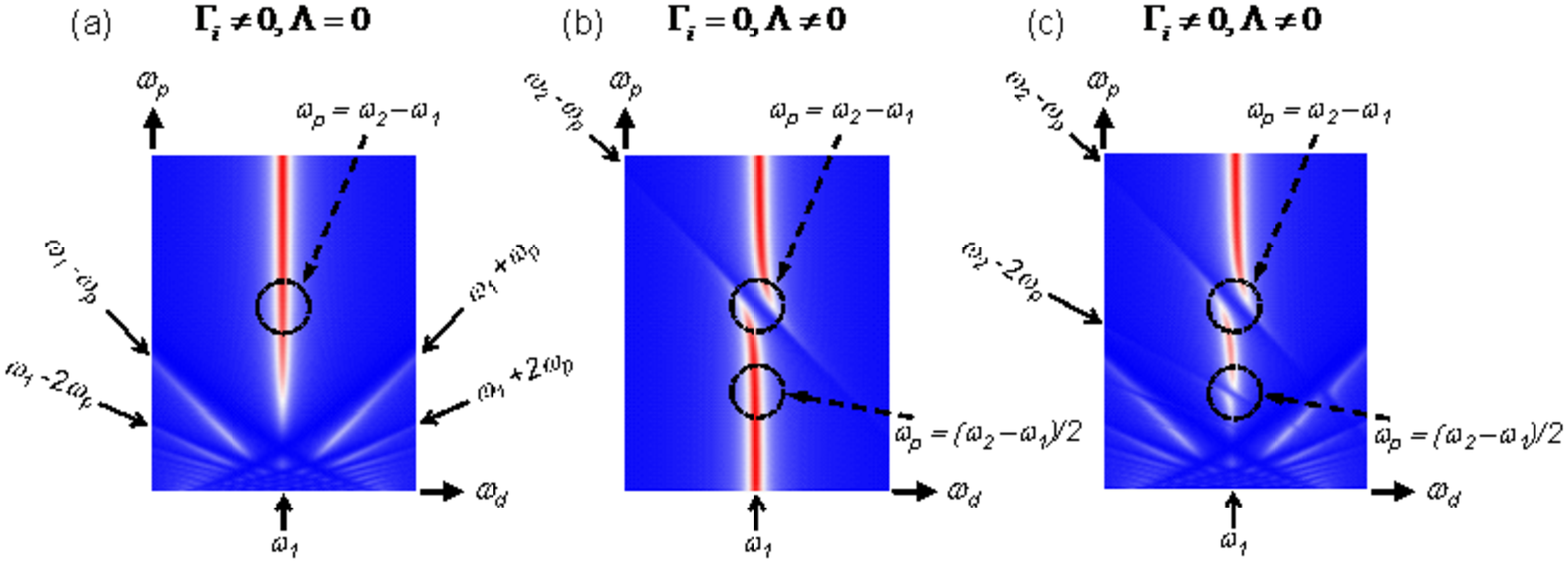}
\caption{Simulated pump frequency dependence of mode 1 by (a) excluding the inter-modal coupling ($\Gamma_i \neq 0$, $\Lambda = 0$), (b) excluding the intra-modal coupling ($\Gamma_i = 0$, $\Lambda \neq 0$), and (c) including both intra- and inter-modal coupling ($\Gamma_i \neq 0$, $\Lambda \neq 0$).}
\label{FigS4}
\end{center}
\end{figure}

\subsection{Inter- and intra-modal coupling}

The role of the inter- and intra-modal coupling can be understood by the simulation for a model, which excludes the inter- or intra-modal coupling terms. When the inter-modal coupling terms are neglected, i.e., $\Gamma_i \neq 0$, $\Lambda = 0$, normal-mode splitting does not occur at $\omega_p \simeq \omega_2 - \omega_1$ (Fig. \ref{FigS4}a). Instead, $\omega_i \pm n\omega_p$ lines appear in the low pump frequency region due to the self-modulation of the $i$-th mode frequency with $\omega_p$, i.e., $\Gamma_i\cos(\omega_p t)x_i$. In contrast, when the intra-modal coupling terms are neglected while the inter-modal coupling is taken into account, i.e., $\Gamma_i = 0$, $\Lambda \neq 0$, normal-mode splitting occurs at $\omega_p \simeq \omega_2 - \omega_1$, whereas $\omega_i \pm n\omega_p$ lines disappear (Fig. \ref{FigS4}b). This reveals that the inter-modal coupling terms including $\Lambda$ plays an essential role in the parametric coupling. It should be noted that the normal-mode splitting appears only at $\omega_p \simeq \omega_2 - \omega_1$ and not at $\omega_p \simeq (\omega_2 - \omega_1)/2$ in this model excluding the intra-modal coupling terms. The normal-mode splitting at $\omega_p \simeq (\omega_2 - \omega_1)/2$ appears only when both the inter- and intra-modal coupling terms are taken into account (Fig. \ref{FigS4}c), indicating that the intra-modal coupling is necessary for the higher-order coupling.

\subsection{Analytical formula for mode splitting}

We derive the analytical formula for mode splitting as a function of pump voltage. We start again from the equations of motion, Eqs. \ref{EqS9a} and \ref{EqS9b}, and perform the perturbation expansion with respect to the pump amplitude, $\Gamma_1$, $\Gamma_2$, and $\Lambda$:
\begin{eqnarray}
& & \ddot{x}_1 + \gamma\dot{x}_1 + \omega_1^2x_1 + (\Gamma_1 x_1 + \Lambda x_2)\cos(\omega_p t) = U_{11}F\cos(\omega_d t + \phi) \nonumber\\
& & \ddot{x}_2 + \gamma\dot{x}_2 + \omega_2^2x_2 + (\Gamma_2 x_2 + \Lambda x_1)\cos(\omega_p t) = U_{21}F\cos(\omega_d t + \phi). \nonumber 
\end{eqnarray}
The above equations can be rewritten as
\begin{eqnarray}
\left( {\bf D_0} + {\bf D_p} \right)
\left( 
\begin{array}{cc}
x_1 \\
x_2 \\
\end{array}
\right) 
= F\cos(\omega_d t + \phi)
\left(
\begin{array}{cc} 
U_{11} \\
U_{21} \\
\end{array}
\right), \label{EqS12}
\end{eqnarray}
where
\begin{eqnarray}
{\bf D_0} =
\left( 
\begin{array}{cc}
\frac{d^2}{dt^2} + \gamma\frac{d}{dt} + \omega_1^2 & 0 \\
0 & \frac{d^2}{dt^2} + \gamma\frac{d}{dt} + \omega_2^2 \\
\end{array}
\right), \;\;\; {\bf D_p} =
\left( 
\begin{array}{cc} 
\Gamma_1\cos(\omega_p t) & \Lambda\cos(\omega_p t) \\
\Lambda\cos(\omega_p t) & \Gamma_2\cos(\omega_p t) \\
\end{array}
\right). \nonumber
\end{eqnarray}
The formal solution for this equation is given by
\begin{eqnarray}
\left(
\begin{array}{cc}
x_1 \\
x_2 \\
\end{array}
\right)
&=& \left( {\bf D_0} + {\bf D_p} \right) ^{-1}F\cos(\omega_d t + \phi)
\left(
\begin{array}{cc}
U_{11} \\
U_{21} \\
\end{array}
\right) \nonumber \\
&=& \left( {\bf D_0}^{-1} - {\bf D_0}^{-1}{\bf D_p}{\bf D_0}^{-1} + {\bf D_0}^{-1}{\bf D_p}{\bf D_0}^{-1}{\bf D_p}{\bf D_0}^{-1} - \cdots \right) F\cos(\omega_d t + \phi)
\left(
\begin{array}{cc}
U_{11} \\
U_{21} \\
\end{array}
\right) \nonumber \\
&=& \sum^{\infty}_{m=0} \left( -1\right) ^m{\bf D_0}^{-1} \left( {\bf D_p}{\bf D_0}^{-1} \right) ^mF\cos(\omega_d t + \phi)
\left(
\begin{array}{cc}
U_{11} \\
U_{21} \\
\end{array}
\right). \label{EqS13}
\end{eqnarray}
The operator ${\bf D_0}^{-1}$ is defined as
\begin{eqnarray}
{\bf D_0}^{-1}
\left(
\begin{array}{cc}
y(t) \\
z(t) \\
\end{array}
\right) =
\left(
\begin{array}{cc}
\left( \frac{d^2}{dt^2} + \gamma\frac{d}{dt} + \omega_1^2 \right) ^{-1}y(t) \\
\left( \frac{d^2}{dt^2} + \gamma\frac{d}{dt} + \omega_2^2 \right) ^{-1}z(t) \\
\end{array}
\right) =
\left(
\begin{array}{cc}
\int\frac{d\omega}{2\pi}\chi_1(\omega)e^{i\omega t}\int dt'e^{-i\omega t'}y(t') \\
\int\frac{d\omega}{2\pi}\chi_2(\omega)e^{i\omega t}\int dt'e^{-i\omega t'}z(t') \\
\end{array}
\right), \label{EqS14}
\end{eqnarray}
where $\chi_1(\omega)$ and $\chi_2(\omega)$ are the mechanical susceptibilities of the two modes:
\begin{eqnarray}
\chi_1(\omega) \equiv \frac{1}{\sigma_1(\omega)} \equiv \frac{1}{-\omega^2 + i\gamma\omega + \omega_1^2}, \;\;\; \chi_2(\omega) \equiv \frac{1}{\sigma_2(\omega)} \equiv \frac{1}{-\omega^2 + i\gamma\omega + \omega_2^2}. \nonumber 
\end{eqnarray}
Using Fourier transformation,
\begin{eqnarray}
\left(
\begin{array}{cc}
\hat{x}_1(\omega) \\
\hat{x}_2(\omega) \\
\end{array}
\right) = \int^\infty_{-\infty}
\left(
\begin{array}{cc}
x_1(t) \\
x_2(t) \\
\end{array}
\right) e^{-i\omega t}dt, \nonumber
\end{eqnarray}
then
\begin{eqnarray}
{\bf \hat{f}(\omega)} &=& \int^\infty_{-\infty}F\cos(\omega_d t + \phi)
\left(
\begin{array}{cc}
U_{11} \\
U_{21} \\
\end{array}
\right) e^{-i\omega t}dt \nonumber \\
&=& \pi e^{i\phi}F\delta(\omega - \omega_d)
\left(
\begin{array}{cc}
U_{11} \\
U_{21} \\
\end{array}
\right) + (\omega_d \rightarrow -\omega_d, \phi \rightarrow -\phi) 
\label{EqS15} 
\end{eqnarray}
\begin{eqnarray}
{\bf \hat{D}_0}^{-1}
\left(
\begin{array}{cc}
\hat{y}(\omega) \\
\hat{z}(\omega) \\
\end{array}
\right) =
\left(
\begin{array}{cc}
\chi_1(\omega)\hat{y}(\omega) \\
\chi_2(\omega)\hat{z}(\omega) \\
\end{array}
\right)
\label{EqS16} 
\end{eqnarray}
\begin{eqnarray}
{\bf \hat{D}_p}
\left(
\begin{array}{cc}
\hat{y}(\omega) \\
\hat{z}(\omega) \\
\end{array}
\right) = \frac{1}{2}
\left(
\begin{array}{cc}
\Gamma_1 & \Lambda \\
\Lambda & \Gamma_2 \\
\end{array}
\right)
\left(
\begin{array}{cc}
\hat{y}(\omega - \omega_p) \\
\hat{z}(\omega - \omega_p) \\
\end{array}
\right) + (\omega_p \rightarrow -\omega_p). 
\label{EqS17}
\end{eqnarray}
The solution can be expanded in the power series of pump operator ${\bf \hat{D}_p}$ as
\begin{eqnarray}
\left(
\begin{array}{cc}
\hat{x}_1(\omega) \\
\hat{x}_2(\omega) \\
\end{array}
\right) &=& \left( {\bf \hat{D}_0} + {\bf \hat{D}_p} \right) ^{-1}{\bf \hat{f}(\omega)} \nonumber \\
&=& {\bf \hat{D}_0}^{-1}{\bf \hat{f}(\omega)} - {\bf \hat{D}_0}^{-1}{\bf \hat{D}_p}{\bf \hat{D}_0}^{-1}{\bf \hat{f}(\omega)} + {\bf \hat{D}_0}^{-1}{\bf \hat{D}_p}{\bf \hat{D}_0}^{-1}{\bf \hat{D}_p}{\bf \hat{D}_0}^{-1}{\bf \hat{f}(\omega)} + \cdots. \label{EqS18}
\end{eqnarray}
The first term provides the $0$-th order approximation. For simplicity, in the following, $\phi$ is assumed to be zero. The unpumped solution is then given by
\begin{eqnarray}
{\bf \hat{D}_0}^{-1}{\bf \hat{f}(\omega)} = \pi F\delta(\omega - \omega_d)
\left(
\begin{array}{cc}
\chi_1(\omega_d)U_{11} \\
\chi_2(\omega_d)U_{21} \\
\end{array}
\right) + (\omega_d \rightarrow -\omega_d). \label{EqS19}
\end{eqnarray}
\begin{figure}[tbp]
\begin{center}
\includegraphics[scale=0.7]{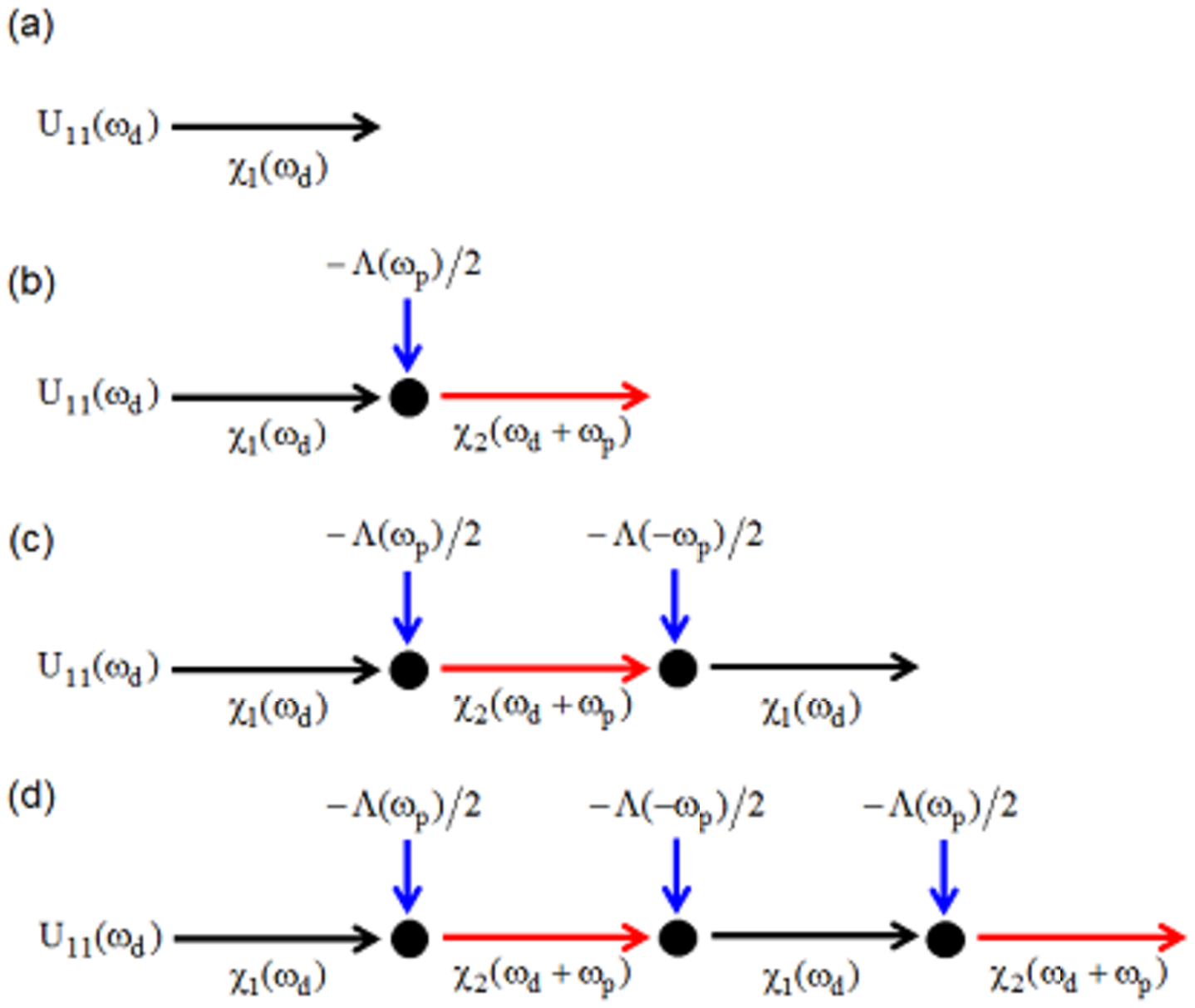}
\caption{Schematic diagram of the (a) 0th-, (b) 1st-, (c) 2nd-, and (d) 3rd-order approximation expressed by Eqs. \ref{EqS20}-\ref{EqS23}, respectively.}
\label{FigS5}
\end{center}
\end{figure}
The solution has a nonzero value only at $\omega = \pm\omega_d$, corresponding to the signal oscillation, and the Lorentz-shape frequency response of the amplitude is given by the susceptibility, $\chi_1(\omega_d)$ or $\chi_2(\omega_d)$. For $\omega_d \sim \omega_1$ and $\omega_2 - \omega_1 >> \gamma$, the contribution from $\chi_2(\omega)$ is negligibly small and only the lower-frequency mode is excited. We can represent this as the diagram shown in Fig. \ref{FigS5}a and approximately obtain
\begin{eqnarray}
{\bf \hat{D}_0}^{-1}{\bf \hat{f}(\omega)} = \pi F\delta(\omega - \omega_d)\chi_1(\omega_d)
\left(
\begin{array}{cc}
U_{11} \\
0 \\
\end{array}
\right) + (\omega_d \rightarrow -\omega_d). \label{EqS20}
\end{eqnarray}
In a similar way, the dominant contribution from the first-order terms is represented as the diagram shown in Fig. \ref{FigS5}b
\begin{eqnarray}
{\bf \hat{D}_0}^{-1}{\bf \hat{D}_p}{\bf \hat{D}_0}^{-1}{\bf \hat{f}(\omega)} = \pi F\delta(\omega - \omega_d - \omega_p)\chi_2(\omega_d + \omega_p)\frac{\Lambda}{2}\chi_1(\omega_d)
\left(
\begin{array}{cc}
0 \\
U_{11} \\
\end{array}
\right) + (\omega_d \rightarrow -\omega_d). \label{EqS21}
\end{eqnarray}
The other term is at off-resonance and the susceptibility is negligibly small so that the term was neglected. The in-coming blue arrow in Fig. \ref{FigS5}b shows that drive frequency $\omega_d$ is up-converted to $\omega_d + \omega_p$ and has a resonance at mode 2. The dominant contribution from the second-order terms is represented as the diagram shown in Fig. \ref{FigS5}c
\begin{eqnarray}
{\bf \hat{D}_0}^{-1}{\bf \hat{D}_p}{\bf \hat{D}_0}^{-1}{\bf \hat{D}_p}{\bf \hat{D}_0}^{-1}{\bf \hat{f}(\omega)} = \pi F\delta(\omega - \omega_d)\chi_1(\omega_d)\frac{\Lambda}{2}\chi_2(\omega_d + \omega_p)\frac{\Lambda}{2}\chi_1(\omega_d)
\left(
\begin{array}{cc}
U_{11} \\
0 \\
\end{array}
\right) & & \nonumber \\
+ (\omega_d \rightarrow -\omega_d).& & \label{EqS22}
\end{eqnarray}
Similarly, the dominant contribution from the third-order terms is represented as the diagram shown in Fig. \ref{FigS5}d
\begin{eqnarray}
& & {\bf \hat{D}_0}^{-1}{\bf \hat{D}_p}{\bf \hat{D}_0}^{-1}{\bf \hat{D}_p}{\bf \hat{D}_0}^{-1}{\bf \hat{D}_p}{\bf \hat{D}_0}^{-1}{\bf \hat{f}(\omega)} = \nonumber \\
& & \;\;\;\;\;\;\;\;\;\;\;\;\;\; \pi F\delta(\omega - \omega_d - \omega_p)\chi_2(\omega_d + \omega_p)\frac{\Lambda}{2}\chi_1(\omega_d)\frac{\Lambda}{2}\chi_2(\omega_d + \omega_p)\frac{\Lambda}{2}\chi_1(\omega_d)
\left(
\begin{array}{cc}
0 \\
U_{11} \\
\end{array}
\right) \nonumber \\
& & \;\;\;\;\;\;\;\;\;\;\;\;\;\;\;\;\;\;\;\;\;\;\;\;\;\;\;\;\;\;\;\;\;\;\;\;\;\;\;\;\;\;\;\;\;\;\;\;\;\;\;\;\;\;\;\;\;\;\;\;\;\;\;\;\;\;\;\;\;\;\;\;\;\;\;\;\;\;\;\;\;\;\;\;\;\;\;\;\;\;\;\;\;\;\; + (\omega_d \rightarrow -\omega_d). \label{EqS23}
\end{eqnarray}
By summing up all the contributions, we finally obtain
\begin{eqnarray}
\left(
\begin{array}{cc}
\hat{x}_1(\omega) \\
\hat{x}_2(\omega) \\
\end{array}
\right) &=& \pi F\delta(\omega - \omega_d)\frac{\chi_1(\omega_d)}{1-\chi_1(\omega_d)\Lambda\chi_2(\omega_d + \omega_p)\Lambda /4}
\left(
\begin{array}{cc}
U_{11} \\
0 \\
\end{array}
\right) \nonumber \\
&-& \pi F\delta(\omega - \omega_d - \omega_p)\frac{\chi_2(\omega_d + \omega_p)\Lambda\chi_1(\omega_d)}{1-\chi_1(\omega_d)\Lambda\chi_2(\omega_d + \omega_p)\Lambda /4}
\left(
\begin{array}{cc}
0 \\
U_{11} \\
\end{array}
\right) + (\omega_d \rightarrow -\omega_d). \label{EqS24}
\end{eqnarray}
\begin{figure}[tbp]
\begin{center}
\includegraphics[scale=0.85]{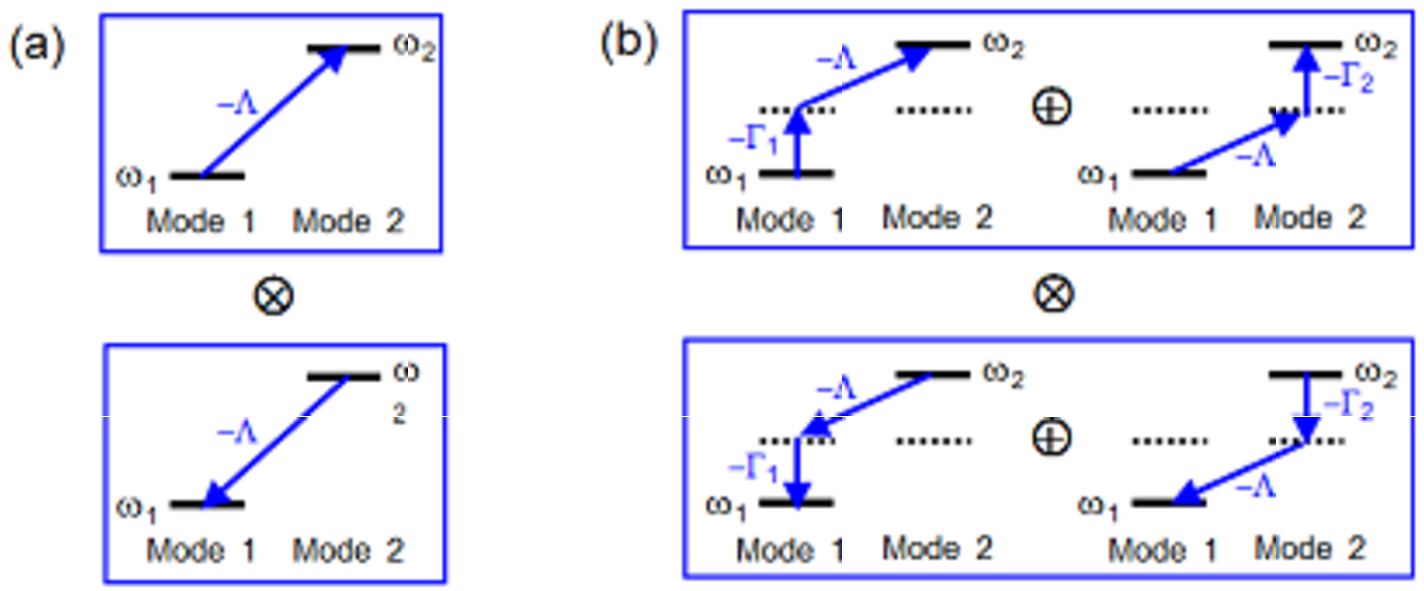}
\caption{Schematic drawing of the (a) one-pump phonon absorption and emission processes and (b) two-pump phonon absorption and emission processes.}
\label{FigS6}
\end{center}
\end{figure}
We now focus on the first term, which gives the renormalized (dressed) spectrum of mode 1. The second term of its denominator is the common ratio of the geometric series and diagrammatically represented as in Fig. \ref{FigS6}a.
For $\omega_p = \omega_2 - \omega_1$ and $\omega_p >> \omega_d - \omega_1 >> \gamma$, we obtain
\begin{eqnarray}
\hat{x}_1(\omega) &=& \pi F\delta(\omega - \omega_d)\frac{\chi_1(\omega_d)U_{11}}{1-\chi_1(\omega_d)\Lambda\chi_2(\omega_d + \omega_p)\Lambda /4} + (\omega_d \rightarrow -\omega_d) \nonumber \\
&=& \pi F\delta(\omega - \omega_d)\frac{\sigma_2(\omega_d + \omega_p)U_{11}}{\sigma_1(\omega_d)\sigma_2(\omega_d + \omega_p) - \Lambda^2/4} + (\omega_d \rightarrow -\omega_d) \nonumber \\
&\sim& \pi F\delta(\omega - \omega_d)\frac{-2\omega_2\delta\omega U_{11}}{(-2\omega_1\delta\omega + i\gamma\omega_1)(-2\omega_2\delta\omega + i\gamma\omega_2) - \Lambda^2/4} + (\omega_d \rightarrow -\omega_d) \nonumber \\
&\sim& \pi F\delta(\omega - \omega_d)\frac{-2\omega_2\delta\omega U_{11}}{4\omega_1\omega_2(\delta\omega)^2 -2i\gamma\omega_1\omega_2\delta\omega - \Lambda^2/4} + (\omega_d \rightarrow -\omega_d).
\label{EqS25}
\end{eqnarray}
In the strong-coupling regime ($\gamma/\delta\omega \rightarrow 0$), it has a maximum when the denominator becomes minimum, i.e., the mode splitting $2\delta\omega^{(1)}$, which corresponds to the coupling rate, $g^{(1)}$, is given by
\begin{eqnarray}
2\delta\omega^{(1)} \sim \frac{\Lambda}{2\sqrt{\omega_1\omega_2}}.
\label{EqS26}
\end{eqnarray}

In the case of two-pump phonon process, i.e., $\omega_p = (\omega_2 - \omega_1)/2$, off-resonance intermediate states at $\omega = \omega_d + \omega_p = \omega_d + (\omega_2 - \omega_1)/2$ should also be taken into account. The diagrammatic representation of the common ratio of geometric series is given by Fig. \ref{FigS6}b. The two dashed lines in Fig. \ref{FigS6}b correspond to the factors $\chi_1(\omega_1 + \omega_p)$ (left) and $\chi_2(\omega_2 - \omega_p)$ (right). They have small values at off-resonance but cannot be neglected for stronger pump amplitude because the transition amplitude is proportional to $V_p^2$. The detailed calculation will be reported elsewhere. The series is given by
\begin{eqnarray}
\hat{x}_1(\omega) = \pi FU_{11}\delta(\omega - \omega_d)\frac{\chi_1(\omega_d)}{1-G^{(2)}} + (\omega_d \rightarrow -\omega_d), 
\label{EqS27}
\end{eqnarray}
where
\begin{eqnarray}
G^{(2)} = \chi_1(\omega_d)\chi_2(\omega_d + 2\omega_p)\left[ \Lambda\chi_1(\omega_d + \omega_p)\Gamma_1 + \Gamma_2\chi_2(\omega_d + \omega_p)\Lambda \right]^2/16. \nonumber
\end{eqnarray}
We can similarly obtained the mode splitting for the second-order process as
\begin{eqnarray}
2\delta\omega^{(2)} &\sim& \frac{| \Lambda\chi_1(\omega_d + \omega_p)\Gamma_1 + \Gamma_2\chi_2(\omega_d + \omega_p)\Lambda |}{4\sqrt{\omega_1\omega_2}}\nonumber \\ 
&\sim& \frac{\Lambda |-\Gamma_1/\omega_1 + \Gamma_2/\omega_2 |}{4(\omega_2-\omega_1)\sqrt{\omega_1\omega_2}}.
\label{EqS28}
\end{eqnarray}

The mode splitting for the third- and fourth-order processes can also be obtained as in the same way to the first- and second-order processes. Here, we only provide their final forms (the detailed calculation will be reported elsewhere):
\begin{eqnarray}
2\delta\omega^{(3)} \sim \frac{9\Lambda |2\Gamma_1^2/\omega_1^2 + 2\Gamma_2^2/\omega_2^2 - (4\Gamma_1\Gamma_2 + \Lambda^2)/\omega_1\omega_2 |}{128(\omega_2 - \omega_1)^2\sqrt{\omega_1\omega_2}}
\label{EqS30}
\end{eqnarray}
\begin{eqnarray}
2\delta\omega^{(4)} \sim \frac{\Lambda |-3\Gamma_1^3/\omega_1^3 + 9\Gamma_1^2\Gamma_2/\omega_1^2\omega_2 + 4\Lambda^2\Gamma_1/\omega_1^2\omega_2 - 9\Gamma_1\Gamma_2^2/\omega_1\omega_2^2 - 4\Lambda^2\Gamma_2/\omega_1\omega_2^2 + 3\Gamma_2^3/\omega_2^3 |}{36(\omega_2-\omega_1)^3\sqrt{\omega_1\omega_2}}
\label{EqS31}
\end{eqnarray}
If we assume that the piezoelectric detuning coefficient as $\alpha/2\Omega_1 = 2\pi \times 0.69$ kHz/V, which is 8\% smaller than that obtained when DC voltage is applied (Fig. \ref{FigS2}a), then we obtain the following theoretical values, which show very good agreement with the experiments as shown in Fig. 4 in the main text, 
\begin{eqnarray}
2\delta\omega^{(1)} \sim 90.06 \; V_{\rm p-p}, \;\;\; 2\delta\omega^{(2)} \sim 60.23 \; V_{\rm p-p}^2, \;\;\; 2\delta\omega^{(3)} \sim 43.19 \; V_{\rm p-p}^3, \;\;\; 2\delta\omega^{(4)} \sim 31.43 \; V_{\rm p-p}^4. \nonumber
\end{eqnarray}
The 8\% difference in the detuning coefficient might be due to the aged shift or the voltage reduction due to the inductance of the electrical connection.

\subsection{Simulations of coherent mechanical oscillations}

\begin{figure}[bp]
\begin{center}
\includegraphics[scale=0.85]{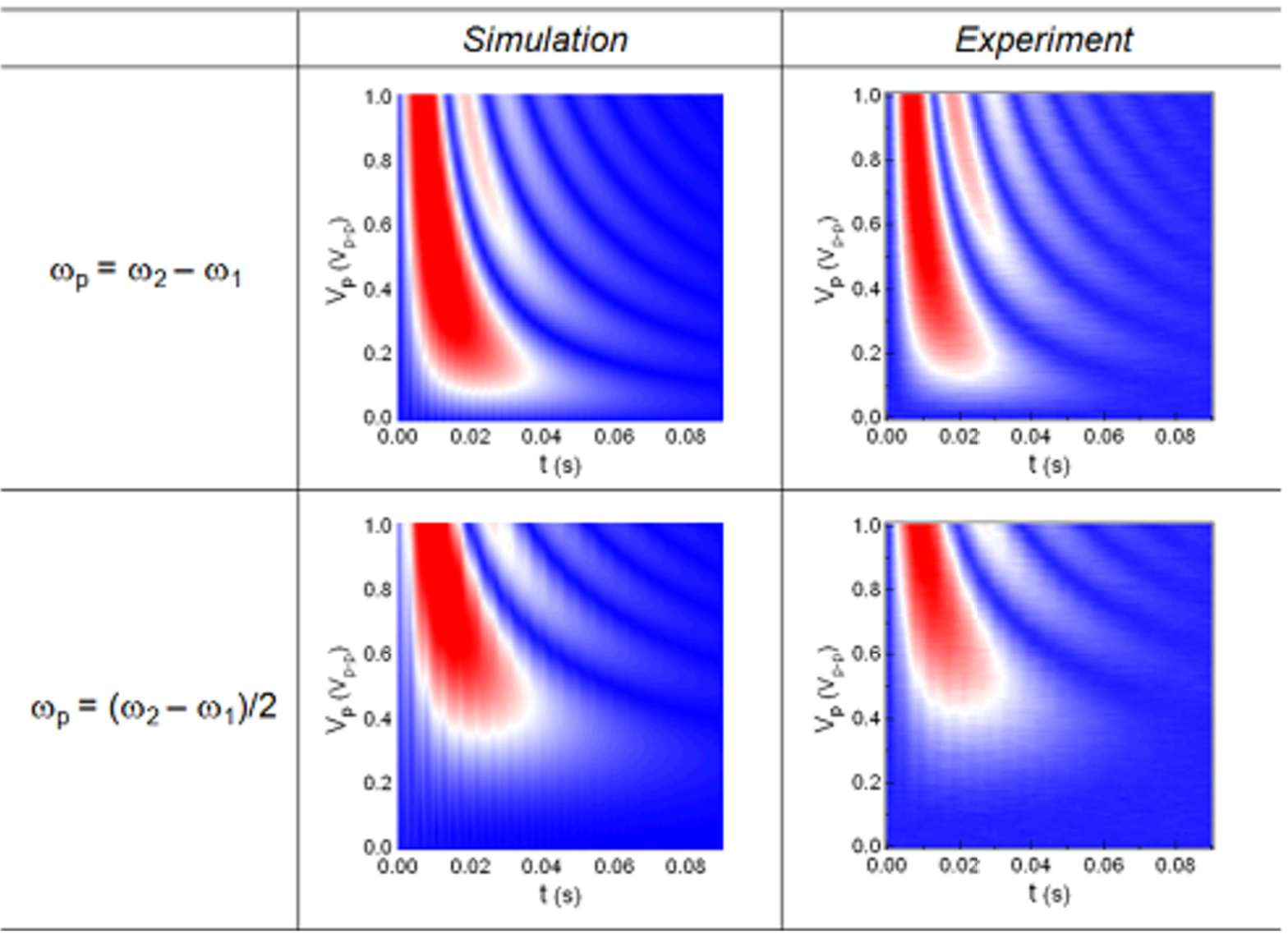}
\caption{Comparison of the pump voltage dependence of the coherent mechanical oscillations between the simulation and the experiment for the first- and second-order coupling.}
\label{FigS7}
\end{center}
\end{figure}

The coherent mechanical oscillations can be simulated by directly solving time-dependent Eqs. \ref{EqS11a} and \ref{EqS11b}. The simulation results for the first- and second-order coupling are shown in Fig. \ref{FigS7} with the experimental results. The experiments show very good agreement with the simulations.

\end{document}